**Angular dependence of magnetoresistivity in c-oriented MgB$_2$ thin film**

C.Ferdeghini, V.Braccini, M.R.Cimberle[b], D.Marré, P.Manfrinetti[c], V.Ferrando, M.Putti, A.Palenzona[c]

INFM, Dipartimento di Fisica, Via Dodecaneso 33, 16146 Genova, Italy

[b]IMEM/CNR, Dipartimento di Fisica, Via Dodecaneso 33, 16146 Genova, Italy

[c]INFM, Dipartimento di Chimica e Chimica Industriale, Via Dodecaneso 31, 16146 Genova, Italy

**Abstract**

The anisotropy of MgB$_2$ is still under debate: its value, strongly dependent on the sample and the measuring method, ranges between 1.1 and 13. In this work we present our results on a MgB$_2$ c-oriented superconducting thin film. To evaluate the anisotropy, we followed two different approaches. Firstly, magnetoresistivity was measured as a function of temperature at selected magnetic fields applied both parallel and perpendicular to the c-axis; secondly, we measured magnetoresistivity at selected temperatures and magnetic fields, varying the angle θ between the magnetic field and the c-axis. The anisotropy estimated from the ratio between the upper critical fields parallel and perpendicular to the c-axis and the one obtained in the framework of the scaling approach within the anisotropic Ginzburg-Landau theory are different. The obtained results are compared and discussed in the light of the two-band nature of MgB$_2$.

**Corresponding author:**

C.Ferdeghini, INFM Dipartimento di Fisica, Via Dodecaneso 33, 16146 Genova, Italy. Tel +39 0103536282, Fax +39 010311066, E-mail: Ferdeghini@fisica.unige.it

**Introduction**

The anisotropy is one of the still not clarified points in the newly discovered [1] superconductor MgB$_2$. Its knowledge is very important in superconducting materials for both basic understanding and practical applications. The anisotropy factor is usually defined as the ratio between the upper critical fields parallel and perpendicular to the basal planes: in the following we will call it $\gamma_{H_{C2}}$. In MgB$_2$ this topic is the object of a vivid debate: the data reported in the literature range between 1.1 and 13, depending on the sample and on the measuring method.

The first approximate estimate of the anisotropy factor was obtained from ac susceptibility measurements on aligned crystallites [2]: $\gamma_{H_{C2}}$ turned out to be 1.7. A surprisingly high value (≈6) was found from conduction electron spin resonance and magnetization measurements performed on powder samples [3, 4]. Successively, measurements on c-oriented thin films appeared in the literature. Patnaik et al. [5] reported anisotropic resistivity from stoichiometric target: they found $\gamma_{H_{C2}}$ values in the range 1.8-2. M.H.Jung et al. [6] estimated a lower value of $\gamma_{H_{C2}} \approx 1.25$ from resistivity measurements on epitaxial film grown from Boron precursor. In [7] we reported $\gamma_{H_{C2}}$ values ranging between 1.8-1.2 in the passage from dirty to clean limit, where we observed also a transition from a temperature independent to a temperature dependent behavior of $\gamma_{H_{C2}}$. Resistivity measurements performed on small single crystals gave values in the interval 2.6-3 [8-10]. More recently, Angst et al. [11] presented torque magnetometry measurements on single crystals and found a temperature dependent anisotropy in the 6-2.8 range with temperature varying from 15 K to $T_C$. Bud'ko et al. confirmed this temperature dependence also by magnetic measurements in powder samples [12]. Sologubenko et al.[13], by analyzing thermal conductivity data on single crystals, found 4 for the anisotropy factor at low temperature, but an upper critical field surprisingly different from the one found resistively both in value and in temperature behavior. The highest anisotropy value $\gamma_{H_{C2}}=13$ was found by Shinde et al. [14] by resistivity measurements performed on low $T_C$ films.

From this overview, it is clear that the issue about the precise anisotropy value is still open: $\gamma_{H_{C2}}$ seems to depend strongly on the kind of sample, on the measurement method, and on the criterion for defining the upper critical field.

In general, thin films, in which the disorder can play a very important role, show less anisotropic behavior: moreover, $\gamma_{H_{C2}}$ seems to depend on the preparation method, the films obtained from boron precursors being less anisotropic than those obtained from stoichiometric ones.In this paper we report on anisotropy of a film grown by a usual two-step technique. To evaluate its anisotropy we followed two different approaches. In the first, following the more usual procedure, magnetoresistivity measurement were performed as a function of temperature at selected magnetic field applied parallel and perpendicular to the c-axis, and $\gamma_{H_{C2}}$ was estimated. In the second, we measured magnetoresistivity at selected temperatures and magnetic fields varying the angle between the magnetic field and the c-axis; thus, by applying a Ginzburg-Landau scaling theory, we obtained the ratio between the in-plane and out-of-plane effective masses, γ, as scaling parameter. We compare and discuss the obtained results and, in the light of the two-band nature of $MgB_2$, we suggest the key role of disorder, strongly enhanced in thin films, in interpreting the great variety of apparently contradictory experimental results on $MgB_2$ anisotropy.

## Sample preparation and characterization

The film was grown by means of Pulsed Laser Ablation by a standard two-step technique. The first step consisted in a room temperature high vacuum deposition of an amorphous precursor layer from $MgB_2$ sintered target [15]. An *ex-situ* annealing in magnesium vapor is needed to crystallize the superconducting phase. Therefore, the sample was placed in a sealed tantalum tube with Mg lumps (approx 0.05 mg/cm$^3$) in Ar atmosphere, and then in an evacuated quartz tube and was heated at T= 850°C for 30 minutes. A rapid quenching to room temperature followed this treatment. The choice of MgO in the (111) orientation was due to its hexagonal surface symmetry, like the one of $MgB_2$, with a lattice mismatch smaller than 3%. X-ray diffraction measurements performed by synchrotron radiation at the ID32 beam line at the ESRF [7] indicated a strong c-axis orientation: in the $\vartheta$-$2\vartheta$ scans both the (001) and (002) reflections of the $MgB_2$ phase were outstanding. Also the peak of the (101) reflection was present, due to a not well oriented fraction of the sample. If we define a texturing coefficient $t = \left[\dfrac{I^{film}(002)}{I^{film}(101)}\right] \cdot \left[\dfrac{I^{powder}(101)}{I^{powder}(002)}\right]$, where $I^{film}(002)$ and $I^{film}(101)$ are the measured intensities of the reflection (002) and (101) and $I^{powder}(002)$ and $I^{powder}(101)$ are the tabulated intensity of randomly oriented powders, we obtain $t \approx 12$, which indicates that only a very small fraction of the film is not oriented. The rocking curve around the (002) reflection, showing a FWHM (Full Width at Half Maximum) of almost 1.3°, confirmed a good c-axis orientation. The sample did not show a single in-plane orientation: however, a strong azimuthal dependence was observed for the (100) reflection in X-rays grazing incidence measurements. The in-plane lattice parameter and the c-axis, calculated from the (100) reflection in grazing incidence measurements and from the (002) reflections measured in symmetrical configuration, turned out to be a=3.073 Å c=3.513 Å, respectively. By comparing these values with those for $MgB_2$ bulk (a=3.086 Å and c=3.524 Å), we remark that the sample is strained: the film adjusts the in-plane lattice with the hexagonal face of the substrate, so reducing the in-plane lattice parameters.

From resistivity measurements we found $T_C$ = 33.7 K, $\Delta T_C$= 1 K, and the residual resistivity ratio RRR=1.5.

## Experimental and discussion

Electrical resistance measurements were performed in a Quantum Design PPMS apparatus in applied magnetic field up to 9T by a four-probe AC resistance technique at 7 Hz. In order to study the anisotropy, we measured the resistance as a function of temperature, magnetic field, and angle

$\theta$ between magnetic field and film surface. The current was always perpendicular to the magnetic field.

From the magnetoresistivity curves as a function of temperature, the upper critical field was estimated in the two directions (with the magnetic field parallel, $H_{C2}(\theta = 0°)$, and perpendicular, $H_{C2}(\theta = 90°)$, to the surface of the sample). The critical field was evaluated at the point of the transition where the resistance is 90% of the normal state value. The results are shown in Fig.1, where the usual phase diagram with $H_{C2}(\theta = 0°)$ values higher than $H_{C2}(\theta = 90°)$ values is reported; the two curves do not decrease linearly to zero but show a positive curvature, more evident in the upper curve ($H_{C2}(\theta = 0°)$). The positive curvature of the upper critical field arises in the clean limit condition [16] and it is magnified in two-band anisotropic systems [17]. This is clearly observable in MgB$_2$ polycrystalline samples and single crystals, while in thin films, as the critical temperature and the residual resistivity ratio RRR decrease, it becomes less pronounced. This behaviour can indicate the passage from clean to dirty limit [7]. The anisotropy factor $\gamma_{H_{C2}} = H_{C2}(\theta = 0°)/H_{C2}(\theta = 90°)$ is plotted in the inset of Fig. 1. $\gamma_{H_{C2}}$ decreases from about 2 to 1.5, as temperature varies from 22 K to T$_C$. A similar temperature behaviour of $\gamma_{H_{C2}}$ has been presented in ref. [11] and [12] and can be inferred also from other upper critical fields data in the literature [5,7,8,10].

Now, the anisotropic behavior of magnetoresistivity measured as a function of the $\theta$ angle at selected temperatures and magnetic fields will be analyzed. The measurements were performed with $\theta$ in the –100° - 100° range, at temperatures of 20, 24, 26, 27.5 and 29 K, and in magnetic fields up to 9 T. As an example, in the insets of Fig.2, two series of data acquired at 24K and 29 K are plotted. As expected, all the curves show a pronounced minimum at $\theta$=0°: as the angle increases (in modulus), resistivity increases and reaches its maximum at $\theta$=±90°. From these curves the $H_{C2}(\theta)$ values can be calculated by drawing a horizontal line corresponding to 90% of the normal state resistivity value: the points where this line meets the magnetoresistivity curves give the $H_{C2}(\theta)$ values directly. In the main panels of Fig.2 the angular dependences of the so calculated $H_{C2}(\theta)$, normalized to $H_{C2}(\theta=90°)$, are reported for T= 24K and 29 K; the values at $\theta$=0 obviously correspond to $\gamma_{H_{C2}}$.

To interpret these experimental data we used the anisotropic Ginzburg-Landau (AGL) theory in which, in the effective mass tensor approximation, the angular dependence of the upper critical field is given by:

$$H_{C2}(\theta) = \frac{H_{C2}(\theta = 90°)}{\left(\sin(\theta)^2 + \frac{1}{\gamma^2}\cos(\theta)^2\right)^{1/2}} \quad (1)$$

where $\gamma = (M_c/M_{ab})^{1/2}$ is the ratio between the effective masses parallel ($M_c$) and perpendicular ($M_{ab}$) to the *c*-axis; it can be straightforwardly derived that $\gamma = H_{C2}(\theta = 0)/H_{C2}(\theta = 90°) = \gamma_{H_{C2}}$. In Fig. 2 eq. (1) is plotted as a continuous line for T=24 K and T=29 K, with the parameter $\gamma$ equal to the previously calculated $\gamma_{H_{C2}}$ (and reported in the inset of Fig.1).

To our knowledge only two papers related to the angular dependence of the upper critical fields appeared in the literature: both set of measurements were performed on single crystals [11][18]. In ref.[18], $H_{C2}(\theta)$, estimated from resistivity measurements, looks very similar to our data, presenting, as in our case, a cusp structure for $\theta = 0$; therefore, these data do not follow eq. (1). The authors underline that this deviation occurs whatever is the criterion chosen for the critical fields definition. On the contrary, in ref. [11] the $H_{C2}(\theta)$ values, calculated from torque magnetometry, follow eq. (1) well for all the angles.

Returning to the data of Fig.2 we note that it is possible to obtain a very good agreement with eq. (1) by neglecting the low angle data using different $\gamma$ values. The best fit curves to the experimental data for $\theta > 20°$ are plotted as dashed lines in Fig. 2. Although there are few data points for each curve, the best fit procedure is accurate, eq. (1) being quite sensitive to $\gamma$ for $\gamma$ values in the range 1-2. On the other hand, the low angle behavior of our data cannot be explained by a not perfect epitaxiality of the film: in fact, un-aligned grains should cause a reduction rather than an increase of the effective anisotropy.

The new $\gamma$ values are reported in the inset of Fig.1 as triangles; they show the same decreasing behavior with temperature as $\gamma_{H_{C2}}$, but they are 20% lower, ranging between 1.8 at 20 K and 1.5 at 29 K. This temperature dependence, as discussed in the following, is certainly out of the AGL schema.

Nevertheless, we try pursuing the analysis of the angular dependence of the magnetoresistivity further within the AGL theory. In this framework Blatter, Geshkenbein and Larkin [19] developed a general scaling approach that makes the treatment of the anisotropic behavior straightforward, at least on a formal basis. Within this model, apart from the region of low dissipation where disorder plays an important role, the resistivity data as a function of angle and magnetic field, if properly scaled, should collapse on the same curve. The rescaled functions are:

$$\tilde{\rho} = \rho \quad (2)$$

$$\tilde{H} = H\left(\sin(\theta)^2 + \frac{1}{\gamma^2}\cos(\theta)^2\right)^{1/2} \qquad (3)$$

We verified the applicability of the scaling rules (2) and (3) on our $\rho(\theta, H)$ data and, for each temperature, we inserted the previously estimated parameter $\gamma$ in $\tilde{H}$. The results are shown in Fig. 3 for T = 29 K: the curves collapse all together and on the measured $\rho(\theta = 90°)$ versus magnetic field curve (continuous line). We mark that this occurs for all the levels of dissipation. Only the low angle data ($\theta < 25°$) do not scale on the main curve, but they bend down with a sharper slope. This failure to scale at low angles, which we observed also in the angular dependence of the upper critical field, is a general feature occurring at all the temperatures. It can be explained within the model [19]: in fact, in a layered superconductor the disorder within the planes and between adjacent planes will not be the same after rescaling, and the difference will be considerable in the small angle regime. This limits the applicability of the scaling to angles such that $|\theta| > arctg(1/\gamma)$. In our case this means $\theta > 25°$-$30°$, this limit becoming less strict for larger anisotropies.

In the inset of Fig. 3 we present $\tilde{\rho}$ as a function of $\tilde{H}$ at T= 20, 24, 26, 27.5 and 29 K, without the low angle data. The scaling is quite good at all the investigated temperatures, even though it becomes less accurate when temperature decreases. Since at lower temperature the resistivity measurements are performed at higher magnetic fields, we cannot distinguish between the influence of the temperature and of the field.

In conclusion, the scaling procedure derived from the AGL approach presents some drawbacks in its application to the case of $MgB_2$. In fact, even though the scaling appears good enough for all the levels of dissipation, temperatures, and fields we tested and in the angular range predicted by the theory, we found an anisotropic factor value that is temperature dependent, in contrast with the theory itself. We recall that in this approach the anisotropy factor is representative of intrinsic quantities, i.e. the effective mass ratio. Therefore, this agreement with the AGL behavior seems to be only formal.

In fact, the non applicability of this theory to $MgB_2$ has been emphasized in [11], where the authors point out that a simple anisotropic mass model cannot take the observed temperature and field dependences of $\gamma$ into account. Thus, following ref. [20], they assume that the anisotropy of the energy gap plays a crucial role in determining the temperature dependence of $\gamma$ in uniaxial superconductors. Actually, unrealistic energy gap anisotropies are required to account for the measured temperature dependence of $\gamma_{H_{C2}}$. Moreover, this model is not able to explain the very different anisotropy values reported in the literature.

In ref. [21] Shulga and co-workers emphasize the necessity of considering a two-band model to account for the upper critical field behavior in MgB$_2$. This model, which has explained the upper critical field behavior of borocarbides [17], introduces many parameters: among them density of state, velocity of electrons, scattering relaxation time with impurity for each band, and coupling between bands. The anisotropy of σ bands observed in MgB$_2$ implements this complex scenario; furthermore, a quantitative analysis of the upper critical field behavior and of its anisotropy has not been performed yet. Anyway, the two-band model can qualitatively explain some features of the upper critical fields behavior. Generally, $H_{C2}(\theta = 0°)$ shows a larger positive curvature than $H_{C2}(\theta = 90°)$, which in some cases is simply linear (the monotone decrease of $\gamma_{H_{C2}}$ with temperature is a consequence of this fact). The two-band model [22] explains both the higher values and the more pronounced curvature of $H_{C2}(\theta = 0°)$ as consequences of the strong anisotropy of the projection of the Fermi surface in the plane θ = 0°.

A second feature is that, in general, single crystals show higher and more temperature dependent anisotropy than films. Differences between bulk and films can be ascribed mainly to the scattering with impurities, which is strongly enhanced in films.

In two-band model, the rising of scattering rate affects the upper critical field behavior [23] in a very complex way. In weakly anisotropic two-band systems, $H_{C2}$ behaves like in single band system, increasing as disorder grows. On the contrary, in strongly anisotropic two-band systems, $H_{C2}$ value decreases and the upward curvature is suppressed as the impurity content grows until the dirty limit is reached; then, a further increase in disorder causes an increase in $H_{C2}$. Because MgB$_2$ is isotropic in-plane and anisotropic out-of-plane, we expect that disorder enhances $H_{C2}(\theta = 90°)$ more than $H_{C2}(\theta = 0°)$. This fact is confirmed by data in the literature where, in thin films, $H_{C2}(\theta = 90°)$ is always higher than in single crystals; on the contrary, $H_{C2}(\theta = 0°)$ does not always assume values higher than in single crystals (even though, in general, it does) and the temperature behavior varies from sample to sample, showing both positive curvature and linear slope. This complex phenomenology explains why in thin films the anisotropy factor $\gamma_{H_{C2}}$ turnes out to be smaller than in single crystals and can be temperature independent: in fact, differences between bulk and films can be ascribed mainly to the scattering with impurities, which is strongly enhanced in films.

**Conclusions**

We have studied the anisotropy of a strongly c-oriented thin film in two different ways: by the ratio between the critical fields perpendicular and parallel to the c-axis and by the angular dependence of resistivity at various fields and temperatures. The AGL theory well accounts for the angular dependence of magnetoresistivity in a large range of temperatures, magnetic fields and angles, and for all the level of dissipation. Nevertheless, a disagreement with the theory, which is based on the effective mass tensor approximation, appears in the temperature dependence of $\gamma$. Moreover, the AGL approach cannot take into account the very different anisotropy values reported in the literature, which seem to be related not to intrinsic properties but mainly to disorder. A qualitative analysis of upper critical fields within a two-band model emphasizes the crucial role of disorder, but a more complex model, taking the anisotropic and multi-band nature of this compound into account, should be developed.

**Figures caption**

Figure 1. $H_{C2}(\theta = 90°)$ (open symbols) and $H_{C2}(\theta = 0°)$ (full symbols) versus temperature. In the inset: $\gamma_{H_{C2}} = H_{C2}(\theta = 0°) / H_{C2}(\theta = 90°)$ (hexagons) as a function of temperature; γ (triangles) obtained from the best fit of $H_{C2}(\theta)$ with eq. (1) (see text).

Figure 2. Main panels: angular dependences of $H_{C2}(\theta) / H_{C2}(\theta = 90°)$ for T= 24 K and T = 29 K; plot of eq. (1) with $\gamma = \gamma_{H_{C2}} = 1.84$ (T = 24 K) and $\gamma = \gamma_{H_{C2}} = 1.63$ (T = 29 K) (continuous lines) and γ = 1.63 (T = 24 K) and γ = 1.5 (T = 29 K) (dashed lines). In the insets: magnetoresistivity as a function of the $\theta$ angle at $\mu_0 H$ = 3, 4, 4.25, 4.5, 4.75, 5, 5.5 T for T= 24 K and at $\mu_0 H$ = 1.2, 1.4, 1.6, 1.8, 2, 2.25, 2.5, 3 T for T = 29 K.

Figure 3. $\tilde{\rho}$ versus $\tilde{H}$ for T = 29 K : the plot shows the $\rho(\theta, H)$ data for ($\mu_0 H$ = 1.2, 1.4, 1.6, 1.8, 2, 2.25, 2.5, 3 T scaled following the relationships (2) and (3); the curves collapse all together on the measured $\rho(\theta = 90°, H)$ curve (continuous line). Inset: $\tilde{\rho}$ versus $\mu_0 \tilde{H}$ for T = 20 K ($\mu_0 H$ = 3, 4, 5, 6, 7, 8, 9 T), T=24 K ($\mu_0 H$ = 3, 4, 4.25, 4.5, 4.75, 5, 5.5 T), T=26 K ($\mu_0 H$ = 2, 3, 3.5, 4, 4.25, 4.5, 4.75 T), T=27.5 K ($\mu_0 H$ = 1.5, 2, 2.5, 2.75, 3, 3.25, 3.5, 3.75, 4 T) and T = 29 K ($\mu_0 H$ = 1.2, 1.4, 1.6, 1.8, 2, 2.25, 2.5, 3 T ) . Only the data for $\theta > 25°$ are plotted.

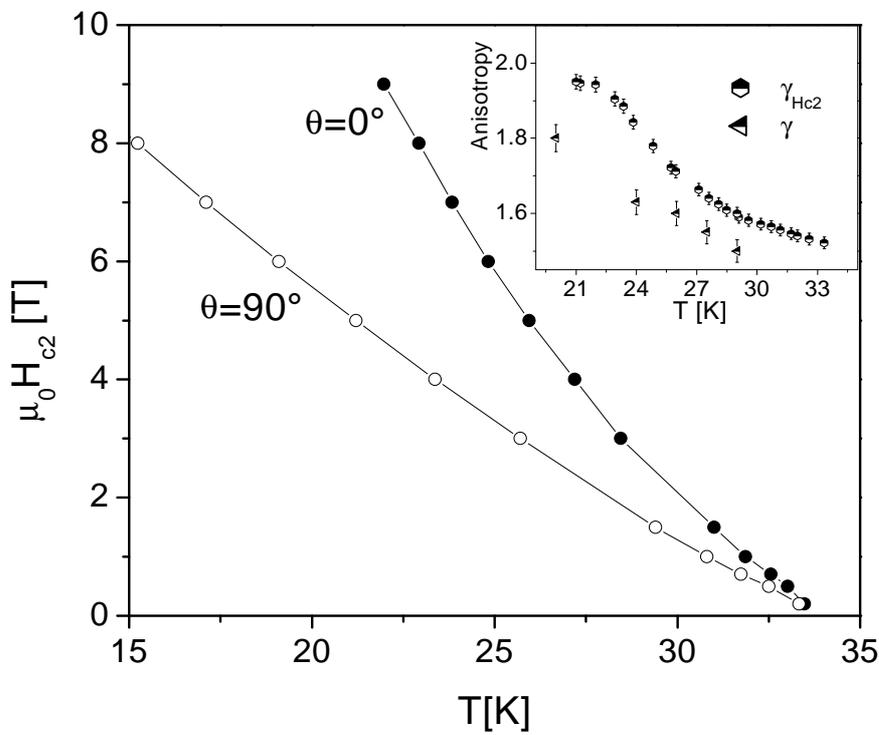

**Figure 1**

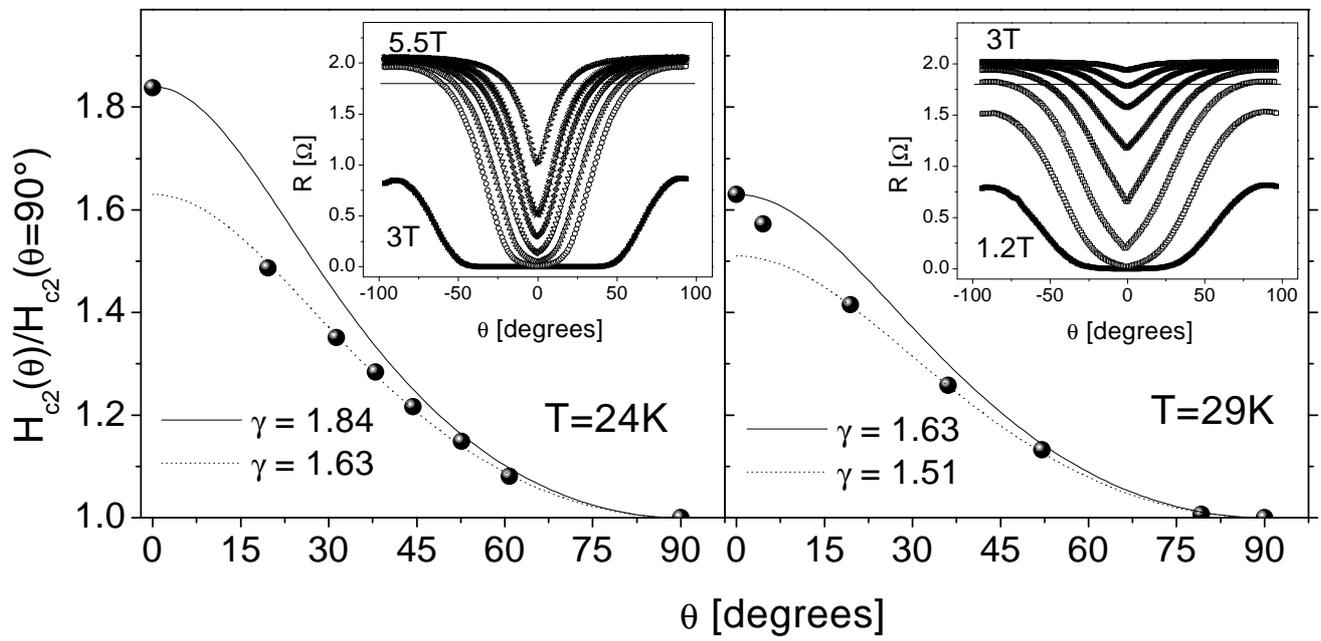

Figure 2

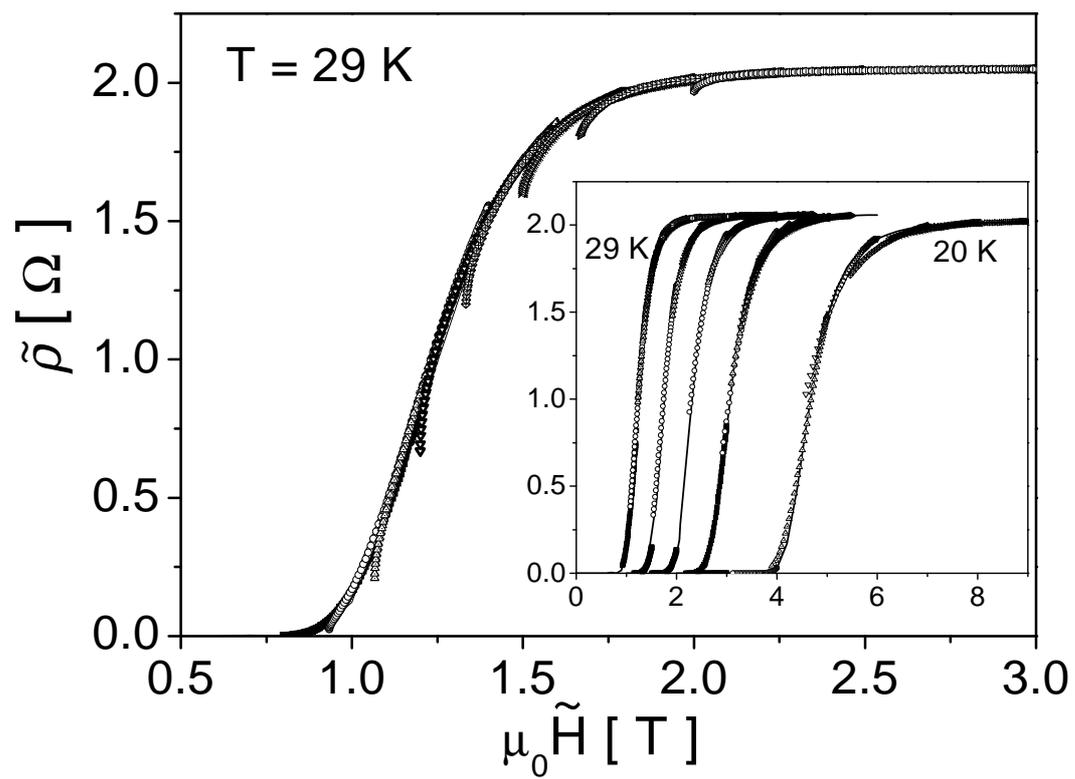

Figure 3